\documentstyle[amssymb,aps,prb]{revtex}

\begin{document}
\title{Distinct charge and spin gaps in underdoped YBa$_{2}$Cu$_{3}$O$_{7-\delta }$
\ from analysis of NMR, neutron scattering, tunneling and quasiparticle
relaxation experiments.}
\author{D.Mihailovic, V.V.Kabanov, K.\v{Z}agar and J.Demsar}
\address{Solid State Physics Dept., ''Jozef Stefan'' Institut, Jamova 39, 1001\\
Ljubljana, Slovenia}
\date{\today}
\maketitle

\begin{abstract}
A systamatic quantitative comparison of ''pseudogap'' values obtained from
the analysis of charge and spin excitation spectroscopies in underdoped YBa$%
_{2}$Cu$_{3}$O$_{7-\delta }$ using a temperature-independent gap shows {\em %
two distinct excitations}, one visible in spin-flip spectroscopies like NMR
and spin polarized neutron scattering, and the other in charge excitation
spectoscopies like single particle tunneling and time-resolved quasiparticle
relaxation. Both appear to decrease with doping $x$ approximately as $1/x$
and are $T$-independent, existing above and below $T_{c}$. We suggest that
the charge excitation can be attributed to a pair-breaking local gap, while
the spin excitation can be explained by an intra-gap local triplet state.
\end{abstract}

\twocolumn The term ''pseudogap'' $\Delta _{p}$ in the context of the
cuprate superconductors usually refers to the observation of a change from
''normal metallic'' behaviour of some measured quantity, which occurs at
some temperature $T^{\ast }$ above the superconducting $T_{c}$. It has been
observed by many different experimental techniques ranging from NMR\cite
{Alloul,Curro,Williams}, specific heat\cite{Loram}, single-particle (SP)
tunneling\cite{Deutscher} to angle-resolved photoemission (ARPES)\cite{ARPES}
and optical conductivity\cite{MMM}. Such an effect can arise if for some
reason a depression in the density of states (DOS) appears, giving rise to a
change in the excitation spectrum, usually a depression in the generalized
susceptibility, single particle spectral density, optical conductivity etc.
However, an alternative view - recently proposed on the basis of femtosecond
quasiparticle relaxation measurements (QPR) in YBa$_{2}$Cu$_{3}$O$_{7-\delta
}$\cite{Kabanov,DemsarPRL} - is that in fact a {\em temperature-independent }%
single-particle (or pair-breaking)\ gap $\Delta _{p}$ exists at all
temperatures in the underdoped state and the appearance of the pseudogap
arises because of a statistical population effect for SP excitations across $%
\Delta _{p}$. When the data were interpreted this way, remarkable agreement
was found between the values of $\Delta _{p}$, with other measurements where
SP\ charge excitations are probed. For example, QPR measurements and SP
tunneling give virtually the same $\Delta _{p}$ in YBCO, while SP\ tunneling
and ARPES also agree quantitatively in Bi$_{2}$Sr$_{2}$CaCu$_{2}$O$_{8}$.
However, the pseudogap $\Delta _{s}$ as observed by NMR\ and spin-polarized
neutron scattering, where spin excitations are probed, appears to be smaller
than the optical and tunneling pseudogap $\Delta _{p}$ for which so far
there is no accepted explanation. Thus in spite of the availability of
spectral data over a large range of doping there is an emerging apparent
discrepancy in the size of the pseudogap.

In this communication, we present a quantitative analysis of the NMR Knight
shift $K_{s}$ in the underdoped state along the lines originally suggested
by Alexandrov and Mott\cite{AM,Alex} for the analysis of the ''spin-gap''
from the temperature-dependence of the NMR relaxation time $T_{1}$. We use
the same type of $T$-independent gap which extends above $T_{c}$ (and $%
T^{\ast })$ as was used to analyze the time-resolved quasiparticle
relaxation data\cite{Kabanov}. The aim is to see if a) the simple model can
describe the $T$-dependence of $K_{s}$ and b) the values of the pseudogap $%
\Delta $ obtained for the NMR $K_{s}$ agree quantitatively with values
obtained from charge excitation spectroscopy and in particular quasiparticle
relaxation measurements, where data for a wide range of doping now exist\cite
{Kabanov,DemsarPRL}. When we compare the results with other experiments, we
find that the charge excitation spectroscopies (QP relaxation, Giever
tunneling and ARPES) all consistently show a pseudogap $\Delta _{p}$ at
approximately twice the size of the spin gap $\Delta _{s}$ (from NMR,
neutron scattering)\ over a wide range of doping. To explain this
observation we propose that the two gaps correspond to two distinct
excitations.

The uniform susceptibility for the case of a pair ground state is given by
the standard equation: 
\begin{equation}
\chi _{s}(T)=-2\mu _{B}^{2}\int_{0}^{\infty }dEN_{s}(E){\frac{\partial f_{s}%
}{{\partial E}}}.
\end{equation}
Here $f_{s}=[y^{-1/2}exp(E/k_{B}T+\Delta _{p}/k_{B}T)+1]^{-1}$ is the
distribution function for electrons (or holes), $y=exp[\mu (T)/k_{B}T]$, $%
\mu (T)$ is the pair chemical potential, $N_{s}(E)$ is the single particle
DOS, $2\Delta _{p}$ is the pair binding energy\ and $\mu _{B}$ the Bohr
magneton. Since the pairing energy (or ''pseudogap'') $\Delta _{p}$ is
expected to be of the order of a few hundred kelvin$,$ we can assume that
single electrons are non-degenerate, and 
\begin{equation}
f_{s}\simeq y^{1/2}\exp \left( -{\frac{E+\Delta _{p}}{k_{B}{T}}}\right) .
\end{equation}
We assume that the single particle DOS can be approximated by: 
\begin{equation}
N_{s}(E)=CE^{\alpha }
\end{equation}
where $C$ is constant, which is in general weakly doping dependent, and $%
\alpha $ is an exponent which characterizes the shape of the single particle
spectrum near the single particle band edge and is typically between -1 and
-1/2. Substituting Eqs.(2) and (3) into Eq.(1) we derive the expression for
the uniform susceptibility: 
\begin{equation}
\chi _{s}(T,x)=C\mu _{B}^{2}y^{1/2}(k_{B}T)^{\alpha }\exp \left( -{\frac{%
\Delta _{p}}{{k}_{B}{T}}}\right) .
\end{equation}
In general, $y$ is a smooth function of temperature, and in the case of high-%
$T_{c}$ superconductors it has been argued\cite{AKM} that $y$ is constant.
Assuming that paired states are non-degenerate above $T_{c},$ we can write
that $y=n_{p}(T)/k_{B}TN_{p}(0)$, where $n_{p}(T)$ is the density of pairs
at temperature $T$ and $N_{p}(0)$ is the density of states, which is assumed
to be constant. Then approximating $n_{p}(T)\sim T$ (which also gives the
correct $1/T$ dependence for the Hall coefficient), we expect that $y$ is
constant over a wide temperature range. (The same temperature independent
behaviour of $y$ is expected if the chemical potential is pinned by a large
DOS in a narrow band.) As a result, Eq. (4) provides a general expression
for the susceptibility of a system with a paired ground state and unpaired
excited state.

From the point of view of the analysis of susceptibility data it is
important to note that formula (4) for a pairing gap is essentially the same
as the susceptibility due to a bound local triplet state above a singlet
paired ground state, namely $\chi _{s}^{S-T}(T)\sim 1/T\exp [-J/k_{B}T]$\cite
{Alex}. Unfortunately this means that a {\em pairing gap cannot be
disitinguished from a spin-flip gap }on the basis of sceptibility
measurements alone.

We can now write a rather general formula for the Knight shift which is
valid irrespective of the origin of the gap and which can apply for {\it %
either }a pairing gap or a spin singlet-triplet gap: 
\begin{equation}
K_{s}=K_{0}+AT^{\alpha }\exp [-\Delta /k_{B}T],
\end{equation}

where $K_{0}$ is the value of $K_{s}$ at zero temperature, and $\Delta $ is
either $\Delta _{p}$ or $\Delta _{s}=J$. (In fact this formula is valid ).

We now proceed by examining the NMR Knight shift data for different nuclei
in YBCO. Fits to the data on YBa$_{2}$Cu$_{4}$O$_{8}$ over a very wide range
of temperatures for $^{63}$Cu by Curro et al\cite{Curro} and up to room
temperature for $^{17}$O by Williams et al\cite{Williams} are shown in
Figure 2 using both $\alpha =-1/2$ and $\alpha =-1.$ Both fits give good
agreement with the data over a very wide range of temperatures; up to 700 K
in the case of $^{63}$Cu. Although the fit with $\alpha =-1/2$ appears to be
slightly better at high temperatures than with $\alpha =-1,$ this should be
treated with caution, since at high temperatures there may be oxygen loss in
the sample, which could probably affect the measurements to some extent. The
choice of $\alpha $ affects the value of $\Delta $ which is obtained from
the fit however, and with $\alpha =-1,$ $\Delta $ can be significantly
smaller than with $\alpha =-1/2,$ as indicated in Fig.1a). Nevertheless, we
conclude that in spite of its simplicity, the model appears to describe the
data very well over a wide range of temperatures, provided the absolute
value of the gap obtained from the fit is treated with caution.

Extending the analysis as a function of doping, the $K_{s}$ data for $^{89}$%
Y NMR\cite{Alloul} on YBa$_{2}$Cu$_{3}$O$_{7-\delta }$ over a wide range of $%
\delta $ are shown in Fig. 1 b)$.$ All the data can be fit with Eq. (5)
using only $\Delta $ as an adjustable parameter. In Fig. 2a) we plot the gap
values $\Delta $ as a function of $\delta $ with both $\alpha =-1/2$ and $%
\alpha =-1$. As before, $\Delta $ appears consistently lower with $\alpha
=-1 $ than with $\alpha =-1/2$ by approximately 100K. Reassuringly, the
values of $\Delta $ obtained from all the different fits for the different
nuclei in Figs. 1a) and 1b) on YBCO-123 and 124 are quite consistent with
each other, which gives us confidence regarding the validity of the approach
to susceptibility analysis using a $T$-independent gap.

In order to interpret the origin of the gap obtained from the analysis of
the NMR\ Knight shift, we now compare these data with other spectroscopies.
In Figure 3 we have plotted the gap as a function of $\delta $ from NMR $%
K_{s}$, spin-polarized neutron scattering (SPNS), tunneling and
quasiparticle relaxation measurements. SPNS shows a spin-excitation peak at $%
E_{s}$=34 meV ($\sim $390 K) in underdoped YBa$_{2}$Cu$_{3}$O$_{6.6}$\cite
{Mook}, which agrees very well with the spin-flip gap from $K_{s}$. On the
other hand, the gap values $\Delta _{p}$ from {\em both }single-particle
tunneling and QPR experiments appear to be significantly larger than $\Delta
,$ by as much as a factor of 2, which strongly suggests that the spin-flip
''gap'' in NMR\ and SPNS, and the quasiparticle gap from QPR\ and tunneling
arise from {\em different excitations}.

This hypothesis is strongly supported by recent neutron scattering data of
Mook et al.\ for $\delta =0.4$ \cite{Mook}. In the neutron scattering phonon
spectra of YBa$_{2}$Cu$_{3}$O$_{6.6}$ there appear to be anomalies which
signify the presence of {\em charge }excitations interfering with phonons.
For $\delta =0.4,$ the anomaly occurs near 70 meV ($\sim $810K), which
agrees remarkably well with the $\Delta _{p}$ from QPR\ and SP tunneling
experiments. Neutron scattering thus appears to show the presence of both
''gaps'', one due to charge excitations and the other due to spin
excitations.

In order to explain the presence of two different ''gaps'' $\Delta _{p}$ and 
$\Delta _{s}$, we start from the original suggestion by Alexandrov and Mott
some years ago for the existence of a triplet state\cite{AM} to propose an
electronic structure of YBaCuO. A schematic diagram is shown in the insert
to Fig. 3. The ground state is composed of {\it local} S=0 singlet pairs.
Since $\Delta _{p}$ is a single-particle charge excitation, it is clearly
associated with pair breaking\cite{Kabanov,Deutscher}. However, if $J<\Delta
_{p}$, the triplet local pair state also exists and lies within the gap. It
should be visible by spin-flip spectroscopies like NMR and SPNS, but not by
optical spectroscopy or SP\ tunneling. We propose therefore that the $\Delta
_{s}$ observed by NMR\ and SPNS is due to a S=1 {\it local pair} triplet
excited state which lies approximately in the middle of the gap. From the
experimental data in Fig. 4, both $\Delta _{p}$ and $\Delta _{s}$ decrease
with increasing doping, more or less as 1/$x$, where $x$ is the hole
density. This is consistent with a reduction of both the pair-binding energy 
$\Delta _{p}$ and the exchange coupling $J$ on the basis of simple screening
arguments\cite{AKM}. For low oxygen doping, at $\delta \sim 0.6,$ $J/k_{B}$ $%
\approx $ 800K, consistent also with 2-magnon Raman scattering measurements 
\cite{Raman}. A rather remarkable feature of these analysis is that a large
body of data on both spin and charge excitation spectroscopies can be
described quantitatively by a very simple model using an isotropic
temperature-independent gap.

We conclude by noting that although we have found that $\Delta _{s}=J\simeq
\Delta _{p}/2$ in YBa$_{2}$Cu$_{3}$O$_{7-\delta }$, the situation might well
be different in other materials. In La$_{2-x}$Sr$_{x}$CuO$_{4}$ for example,
the optical gap and the NMR\ gap appear to have the same energy scale\cite
{Mueller}, suggesting that either $J\gtrsim \Delta _{p},$ or that the
triplet pair state is not a bound state. It will be interesting to see if
analysis of other materials might lead to some systematic connection between 
$\Delta _{p,}$ $J$ and $T_{c}$.

%
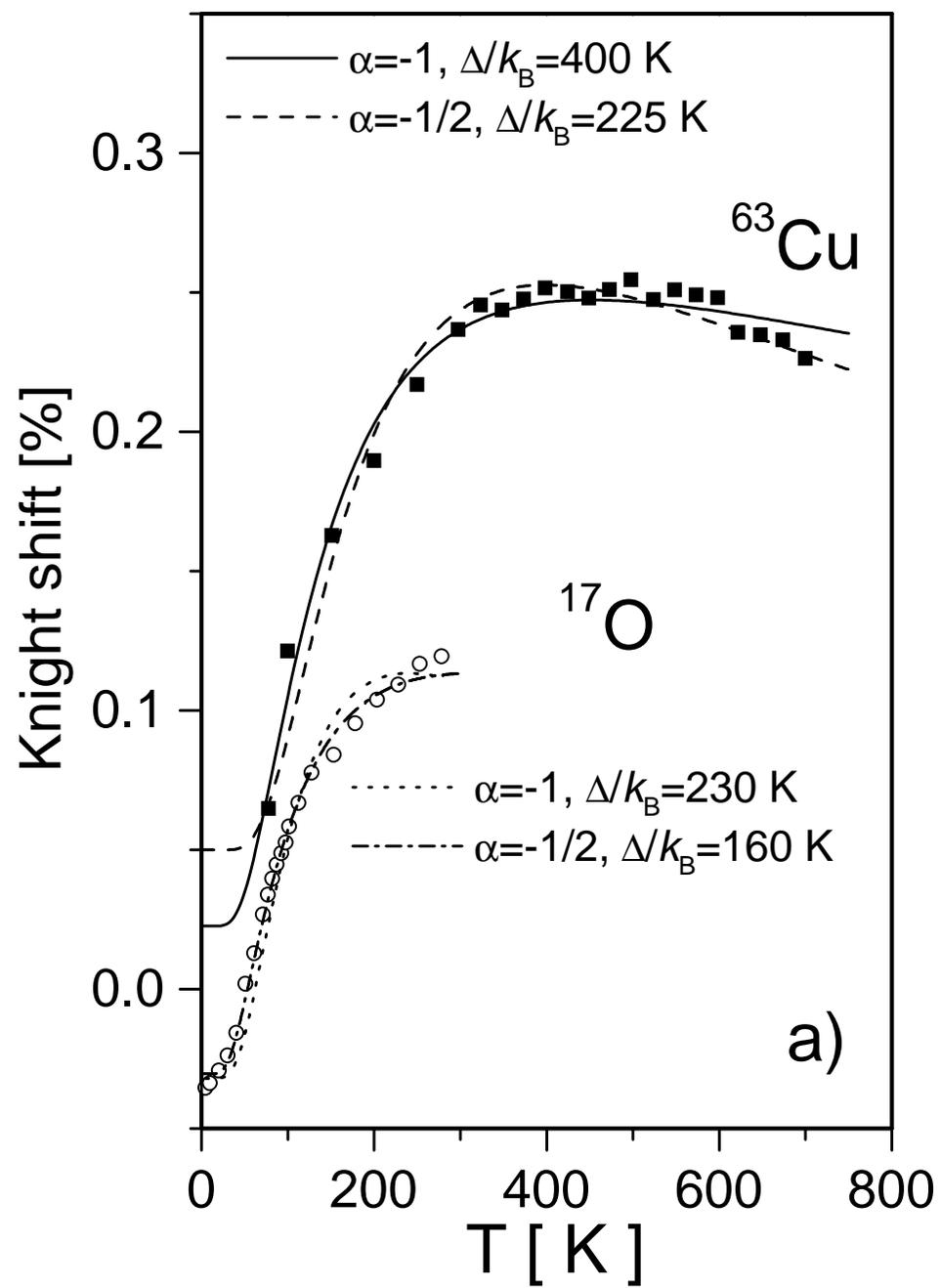
\begin{figure}[hb]
\caption{The 
Knight shift K$_{s}$ as a function of temperature 
a) in YBa$_{2}$Cu$_{4}$O$_{8}$ ($T_{c}$=81 K). 
The
solid squares are for $^{63}$Cu from Curro et al \protect\cite{Curro}
and the open circles for $^{17}$O from Williams et al \protect\cite{Williams}. The parameters 
used in the fits with Eq. 5 are given in the figure.
b) The $^{89}$Y NMR  K$_{s}$ as a function of temperature in YBa$_{2}$Cu$_{3}$O$_{7-\delta }$ 
for different $\delta $\protect\cite{Alloul}, The fits with $\alpha =-1$ and $\alpha=-1/2$
are indistinguishable on the plot, but give different values of $\Delta$.}
\end{figure}

%

%
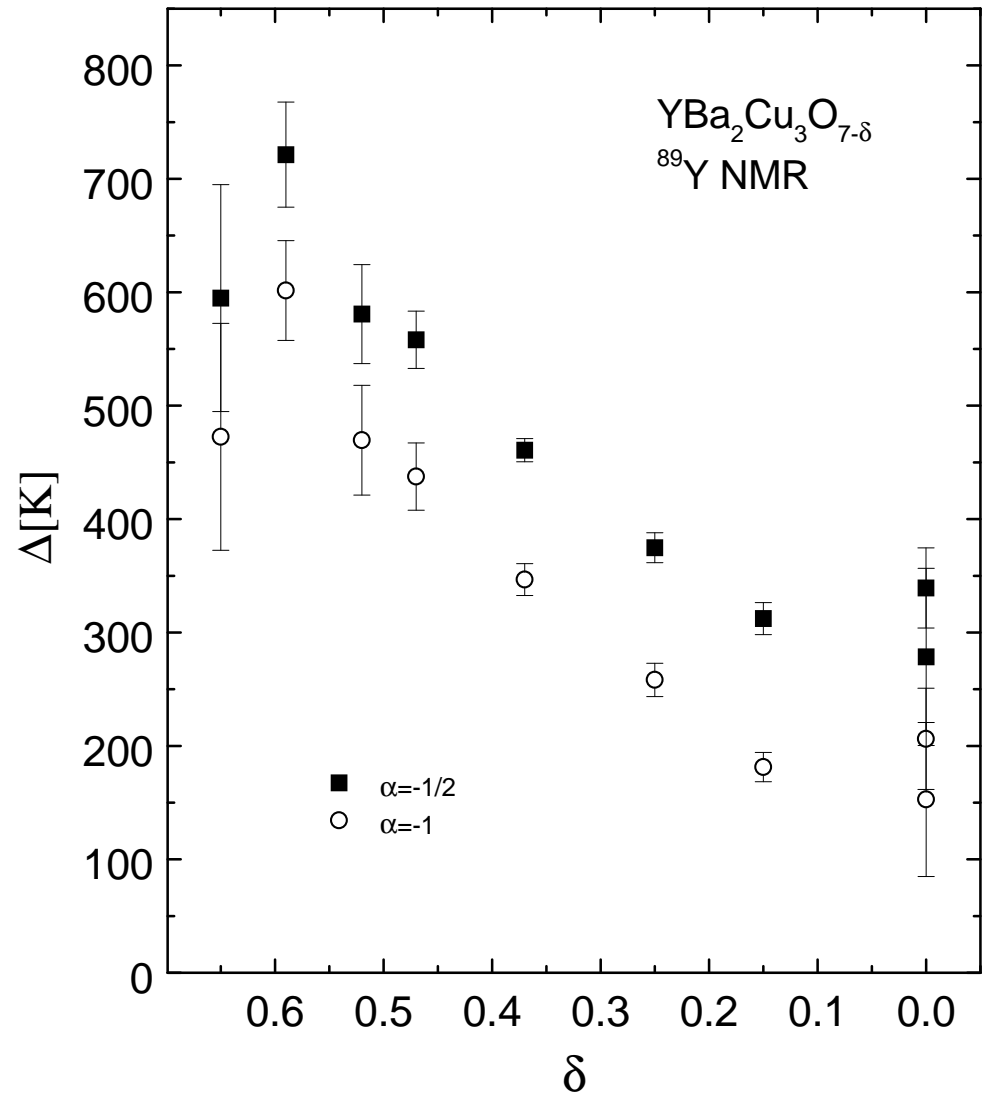
\begin{figure}[hb]
\caption{
The energy gap(s) $\Delta$ as\ a function of doping $\delta$. The circles are for 
$\alpha=-1$ and the squares for $\alpha=-1/2$.
}
\end{figure}

%

%
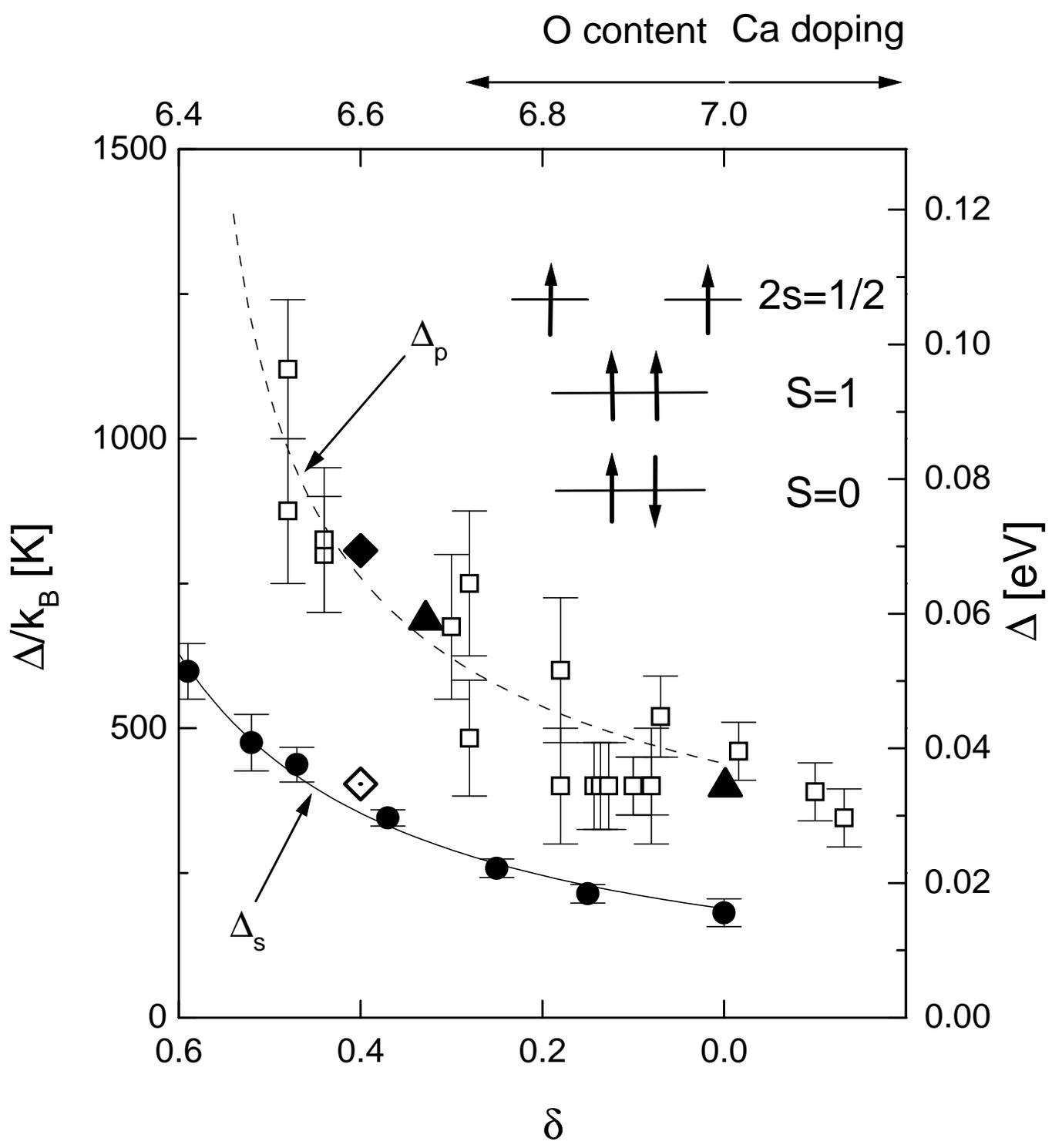
\begin{figure}[hb]
\caption{The charge and spin gaps $\Delta _{p}$ and $\Delta _{s}$ respectively, as\ a function of doping. 
The full circles are from $^{89}$Y NMR $K_{s}$. 
The open squares are from
time-resolved QP relaxation measurements\protect\cite{Kabanov,DemsarPRL}. 
The open and full diamonds are the SPNS and charge excitation 
neutron data respectively, while the triangles are from tunneling
data \cite{Deutscher}.
}
\end{figure}

%

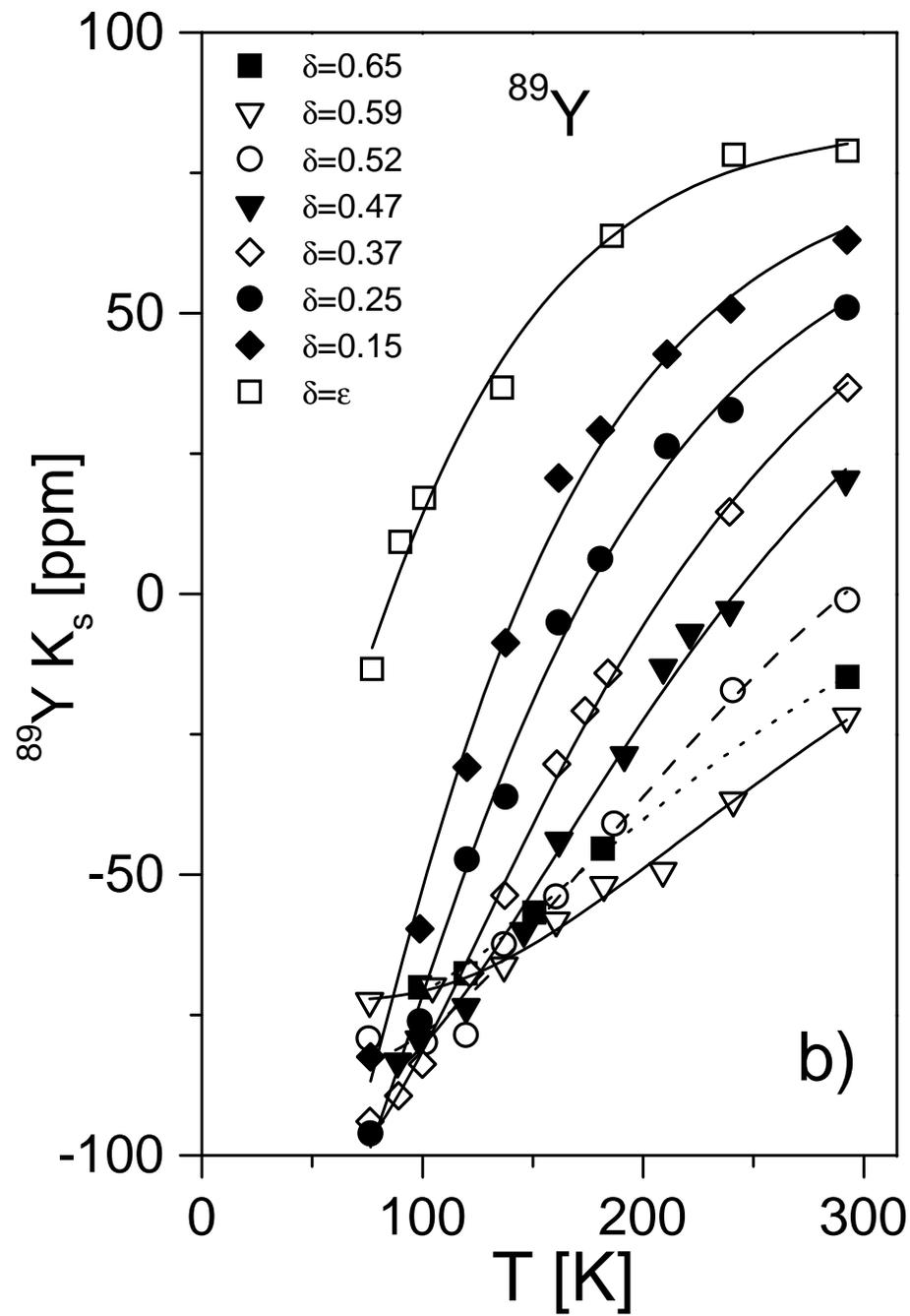
\end{document}